\documentclass[a4paper,fleqn]{cas-sc}

\usepackage[numbers,sort&compress]{natbib}

\usepackage{amsmath,amssymb,amsfonts}
\usepackage{booktabs}
\usepackage{multirow}
\usepackage{url}
\usepackage[most]{tcolorbox}
\usepackage[noend]{algorithmic}
\usepackage[ruled,vlined]{algorithm2e}
\usepackage{enumitem}
\usepackage{alltt}
\usepackage{amsthm}
\usepackage{mathtools}
\usepackage{subcaption}
\usepackage{soul}
\usepackage{float}
\usepackage[section]{placeins} 
\usepackage{listings}

\graphicspath{{Figures/}}

\tcbset{
  aibox/.style={
    width=\textwidth,
    top=10pt,
    colback=white,
    colframe=black,
    colbacktitle=black,
    enhanced,
    center,
    attach boxed title to top left={yshift=-0.1in,xshift=0.15in},
    boxed title style={boxrule=0pt,colframe=white,},
  }
}
\newtcolorbox{AIbox}[2][]{aibox,title=#2,#1}

\definecolor{aigold}{RGB}{244,210, 1}
\definecolor{aigreen}{RGB}{210,244,211}

\sethlcolor{aigreen}

\definecolor{aired}{RGB}{255,180,181}

\newtcbox{\mybox}[1][green]{on line,
arc=0pt,outer arc=0pt,colback=#1!10!white,colframe=#1!50!black,
boxsep=0pt,left=0pt,right=0pt,top=0pt,bottom=0pt,
boxrule=0pt,bottomrule=0pt,toprule=0pt}

\newcolumntype{P}[1]{>{\centering\arraybackslash}p{#1}}

\begin{document}
\let\WriteBookmarks\relax
\def\textfraction{0.1}
\def\topfraction{0.9}
\def\bottomfraction{0.8}
\def\floatpagefraction{0.7}
\setcounter{topnumber}{9}
\setcounter{bottomnumber}{9}
\setcounter{totalnumber}{20}
\setcounter{dbltopnumber}{9}

\shorttitle{LLM-Powered Agentic Workflow for Power System TSA}

\shortauthors{Hu et~al.}

\title [mode = title]{LLM-Driven Transient Stability Assessment: From Automated Simulation to Neural Architecture Design}

\author[a]{Lianzhe Hu}
\ead{hulianzhe@163.com}
\credit{Conceptualization, Methodology, Software, Writing - Original draft}

\affiliation[a]{organization={School of Electrical Engineering, Chongqing University},
            addressline={Shazheng Street},
            city={Chongqing},
            postcode={400044},
            state={},
            country={China}}

\author[a]{Yu Wang}
\ead{yu_wang@cqu.edu.cn}
\credit{Conceptualization, Supervision, Funding acquisition, Validation, Writing - Review \& editing}

\author[b]{Bikash Pal}
\ead{b.pal@imperial.ac.uk}
\credit{Supervision, Writing - Review \& editing}

\affiliation[b]{organization={Department of Electrical and Electronic Engineering, Imperial College London},
            addressline={South Kensington Campus},
            city={London},
            postcode={SW7 2AZ},
            state={},
            country={United Kingdom}}

\cortext[1]{This work is supported by the National Natural Science Foundation of China. Grant No.: 52577084. (Corresponding author: Yu Wang, email: yu\_wang@cqu.edu.cn).}

\begin{abstract}
Existing transient stability assessment (TSA) methods rely heavily on manual simulation-based data generation and hand-crafted neural architectures, which limits the efficiency of the workflow. This paper proposes an LLM-driven agentic framework that automates the complete TSA workflow from simulation to neural network model design. Through integration LLM-agent with the ANDES simulator, natural language instructions are translated into executable disturbance scenarios without manual scripting. Specifically, domain knowledge and simulation templates are encoded into structured prompts with Retrieval-Augmented Generation (RAG) to enhance reasoning via Chain-of-Thought (CoT). Besides, an LLM-driven neural network design pipeline automatically discovers optimized architectures through iterative performance feedback, replacing traditional manual architecture engineering. To ensure reliable automation, a self-correction mechanism collects runtime errors and validation metrics for continuous refinement. Case studies on the IEEE 39-bus system and IEEE 118-bus system demonstrate that the proposed approach achieves superior accuracy and computational efficiency compared to manually-designed baselines. Ablation studies confirm that retrieval augmentation, reasoning enhancement, and feedback mechanisms jointly contribute to robust performance. The results validate effective LLM-driven automation for TSA and encourage further exploration in broader power system analysis tasks.
\end{abstract}

\begin{highlights}
\item End-to-end agentic LLM workflow automates TSA from simulation to model design
\item Natural language--driven scenario generation via tight LLM--ANDES integration
\item LLM--NND discovers compact models (4.78M params) with 93.71\% accuracy
\item Paradigm readily extends to other power system analysis tasks
\end{highlights}


\maketitle

\section{Introduction}\label{sec:introduction}
Modern power grids are undergoing a fundamental transformation driven by the global energy transition, characterized by the increasing integration of high-penetration renewable energy sources (RES) and power-electronic-interfaced devices \citep{kundur1994power,milano2018lowinertia}. This shift introduces unprecedented complexities in network topology, dynamic characteristics, and stability mechanisms. Consequently, this transformation, while critical for a sustainable energy future, poses severe challenges to conventional TSA methodologies, which were designed for traditional, centralized generation systems. The urgency of this issue is underscored by the frequent occurrence of related power system security incidents. Notable examples include the 2016 South Australia blackout, attributed to high wind power integration \citep{yan2018southausblackout,aemo2017southausreport}, and the 2025 Iberian Peninsula blackout \citep{expertpanel2025iberian}, where an initial small-scale generation tripping event—against a backdrop of high renewable energy penetration—rapidly escalated into a large-scale cascading failure.

The transient stability of a power system refers to its ability to maintain synchronous operation after being subjected to significant disturbances. Traditional power system stability assessment generally relies on mathematical and physical methods, such as time-domain simulation, energy function method \citep{fouad1991tef,pai1989energy}, and analytical calculation. Although these methods can provide interpretable results, in modern power systems, they have very prominent limitations. The computational complexity of them will grow rapidly as the system scale increases and renewable energy is added. Scenario generation requires a lot of expert knowledge and human intervention. The process of model design and optimization largely relies on domain expertise, which limits its scalability and makes it difficult to adapt to the constantly changing grid conditions. Nowadays, a large number of new power electronic devices are integrated, and various grid following or forming control strategies \citep{rocabert2012microgridcontrol} are applied. This has increased the demand for analyzing multi-time-scale characteristics. The proportion of new energy usage is constantly rising, and flexible high-voltage direct current transmission \citep{flourentzou2009vschvdc} and other technologies have also begun to be used. This has made the grid architecture increasingly complex and larger in scale. The connections among the various parts have also become closer. The traditional calculation methods are very time-consuming. The system is too complex, which also makes it difficult to calculate or represent the energy function, thus resulting in a large amount of computational work.

Meanwhile, with the rapid development of artificial intelligence theory and computing hardware, the technical route of conducting stability analysis based on artificial intelligence technology has gradually become obvious \citep{zhang2021critical}. The model architecture has become increasingly complex. At first, it was shallow machine learning, and later it developed to complex deep neural networks \citep{zhang2021critical}, and its functions are also constantly evolving. Initially, the task was framed as a straightforward binary classification problem. Later, it could provide stability margin assessment and also support security-oriented decision-making \citep{zhang2021critical}. The power system stability assessment technology using artificial intelligence has attracted extensive attention from researchers due to its high efficiency and accuracy. To complete this task, many machine learning models have been proposed. Large language models are a very crucial branch in artificial intelligence. Its realization relies on algorithmic innovation in natural language processing, advancements in computing power, and the diffusion of big data \citep{vaswani2017attention,brown2020language,devlin2019bert}. These technological breakthroughs have significantly enhanced the ability of artificial intelligence systems to absorb knowledge, decompose complex tasks, and handle multimodal data \citep{wei2022chain}. By integrating LLMs and professional knowledge in the field of power systems within a modular framework, these models can independently and accurately complete a wide variety of simulation tasks, even in previously unseen simulation environments \citep{jia2024daline}. There is a feedback-driven multi-agent framework that can significantly enhance the task success rate of LLMs on simulation platforms \citep{jia2024daline}. Technologies such as RAG, prompt engineering, and CoT can integrate power system knowledge into LLMs \citep{liu2023pretrain,wei2022chain}. The large-scale "Grid-LLM" model trained on power system data and documents in specific domains provides a new way to obtain simulation data \citep{majumder2024exploring}. Well-trained LLMs can make time series predictions on multi-dimensional data such as load, electricity price and renewable energy data. LLMs can also directly solve simplified optimal power flow problems through natural language interaction. Combining grid graph embedding technology with LLMs has also been proven to achieve efficient OPF solutions \citep{yan2023realtime,bernier2025powergraphllm}. These studies all indicate that LLMs have great potential in a more comprehensive stability assessment workflow.

Although the application of large language models in the field of power systems has developed quite rapidly, up to now, their application in the stability assessment of power systems is still rather scattered and the application scope is also relatively limited. The existing methods generally only handle individual parts of the assessment workflow, such as data preprocessing, feature extraction or result interpretation, etc. It is impossible to form a unified overall framework, and a large number of people still need to make adjustments. Although neural architecture search has brought significant changes to the design of automated models in fields such as computer vision and natural language processing \citep{elsken2019neural,ren2021comprehensive,zoph2016nas}, its application in the stability assessment of power systems is still relatively limited. Currently, the deep learning methods used for transient stability assessment rely on artificially constructed architectures. This limits their adaptability to various power system configurations and operating conditions. This article has conceived a model shift from using manually designed, fragmented tools to agentic workflows driven by large language models. If properly arranged, large language models can act as multiple agents to manage the entire process, which includes data simulation for stability studies and the construction of neural network models for transient stability assessment.

Building upon the aforementioned concepts and exploring the feasibility of LLM-driven applications in TSA, this paper makes three primary contributions by integrating LLMs as intelligent agents into a complete power system TSA workflow:

\textbf{LLM-Driven Simulation Control Agent:} We first propose a novel LLM-simulator integration framework that translates natural language descriptions into executable simulation scenarios. By leveraging prompt engineering and RAG to enhance the LLM's domain-specific understanding of transient simulations, this agent achieves autonomous scenario generation, automated parameter configuration, and LLM-based result interpretation, while simultaneously broadening the coverage of critical operating conditions.

\textbf{LLM-Driven Neural Network Design Agent:} We then propose a framework that employs an LLM to conduct Neural Architecture Search (NAS). This agent autonomously designs, optimizes, and validates neural network architectures via a performance-driven feedback loop, iterating until the TSA efficacy meets the specified requirements.

\textbf{Framework Validation and Efficacy Assessment:} Finally, we conduct comprehensive case studies to validate the feasibility and practical effectiveness of the proposed framework. This validation involves a comparative analysis of transient simulation convergence and an assessment of the performance metrics of the neural network architectures designed by the LLM agent.

This paper provides a comprehensive documentation of our framework and its validation. Section 2 details the methodology of each pillar, including the technical architecture, implementation strategies, and theoretical underpinnings. Section 3 presents the comprehensive experimental validation, featuring performance benchmarks, ablation studies, and a case study analysis that highlights the advantages of the LLM-driven approach. Finally, Section 4 concludes with strategic insights for future research directions and discusses the broader implications for AI-driven scientific discovery in critical infrastructure applications.

\section{LLM-Powered Agentic Workflow for Power System TSA}\label{sec:methodology}

This section describes the proposed framework for automating the TSA workflow from simulation to model design. The framework consists of two core modules: the LLM-Driven Simulation Controller, which converts natural language instructions into executable simulation scenarios, and the LLM-Driven Neural Network Designer (LLM-NND), which automatically searches for optimal architectures through iterative feedback. Both modules incorporate RAG for domain knowledge and CoT prompting with self-correction to improve reliability. The technical details of each module are presented below.

\subsection{LLM-Driven Simulation Controller}

\begin{center}
\includegraphics[width=\textwidth]{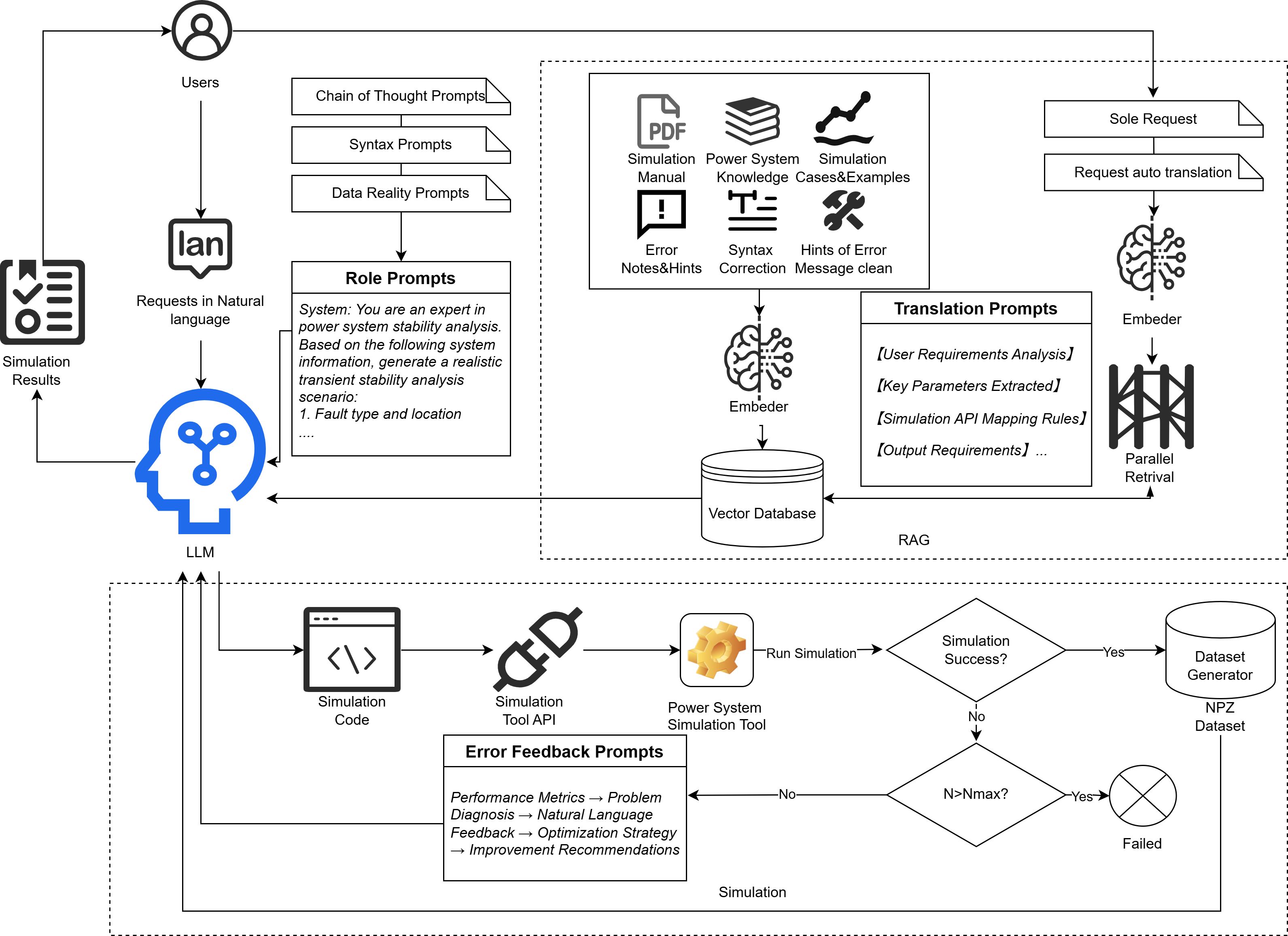}
\captionof{figure}{Overall architecture of LLM-driven simulation controller framework.}
\label{fig:framework1}
\end{center}

\vspace{0.5cm}

Traditional transient stability assessment of power systems largely relies on expert knowledge in scenario design and parameter configuration, which takes a lot of time and may not be able to include all key operating conditions \citep{pavella2012transient}. To address these existing limitations, a brand-new framework has been developed, which utilizes large language models as simulation controllers. This is to achieve automated scenario generation and data collection.

The method we adopt is fundamentally different from traditional simulation methods. It introduces natural language understanding capabilities into the simulation workflow. Large language models can act as intelligent intermediaries between human operators and simulation tools. They can interpret advanced requirements and transform these requirements into executable simulation configurations. This shift in the model has made the stability assessment process more efficient and comprehensive.

Figure 1 shows the overall architecture of LLM-driven simulation controller framework. The proposed framework consists of four modules and multiple technologies: (i) Prompt Engineering, (ii) Enhanced RAG, and (iii) Simulation and Feedback Loop, as illustrated in Figure 1 and detailed below. In this work, we use the ANDES simulator \citep{cui2018andes,cui2025andesmanual} as the simulation backend.

\subsubsection{Prompt Engineering}

Prompt engineering is not merely about designing and developing prompts; it actually encompasses a wide range of skills and techniques for interacting with large language models, or LLMs. This has been mentioned in relevant literature \citep{white2024promptpattern,naser2025prompt}. Prompt engineering plays a significant role in integrating with large language models and understanding their capabilities. It plays a very crucial role. A general prompt usually contains any one of the following elements, as described in Reference \citep{liu2023pretrain}. The first element is the instruction, which refers to the specific task or instruction that the model needs to perform. The second element is the context, which is the external information or additional circumstances that can guide the model to respond better. The third element is the input data, which refers to the content provided by the user or the questions raised. The fourth element is the output metric, which is used to specify the type or format of the output. With these elements, the model can be guided more effectively, and thus better results can be produced.

Under the current background where power system simulation can be rapidly designed, most of the related research currently carried out comes from the field of large language models. The main focus of these studies is to help LLMs understand the role they play in this process and what they are researching. In this way, they can use simulation tools to carry out relevant analysis work. In order to formulate a relatively comprehensive prompt to guide large language models to fully exert their role in power system simulation, the CoT prompt approach can be adopted. By means of step-by-step intermediate steps, its ability in complex reasoning can be enhanced \citep{wei2022chain,kojima2022zeroshot}. Some existing studies also utilize a small number of prompt methods to achieve context learning \citep{brown2020language,dong2024icl}, that is, providing some demonstrations in the prompt content to guide the model to achieve better performance.

Beyond referencing these established strategies, we have designed a 'step-by-step' prompt—verified to elicit CoT reasoning—to enhance the LLM's self-awareness of its function. This process operates sequentially: first, a prompt ensures that CoT is enabled for simulation identification; this is followed by a prompt that employs an Automatic Prompt Engineer (APE) for automatic syntax optimization\citep{zhou2022humanlevelprompt}; subsequently, a 'simulation factuality' prompt prevents the LLM from fabricating data due to hallucinations; finally, a prompt facilitates the LLM's understanding and conversion of natural language. By integrating these techniques, all proven effective in the LLM domain, we define a foundational, comprehensive prompt that establishes the LLM's awareness of its functional role. These prompts are infused with domain-specific knowledge and experience from power system simulation to improve their effectiveness.

\begin{tcolorbox}[top=10pt, colback=white, colframe=black, colbacktitle=black, center, enhanced, unbreakable,
attach boxed title to top left={yshift=-0.1in,xshift=0.15in}, boxed title style={boxrule=0pt,colframe=white,}, title=Example Prompt for Scenario Generation]
{\bf Role Prompt:} ``You are an expert in power system transient stability assessment. Your task is to generate realistic fault scenarios for a power system with the following characteristics: [system description]. Based on the system topology, load distribution, and critical operating points, propose fault scenarios that comprehensively test the system's stability.''

{\bf Architecture Prompt:} ``Generate a Python script using ANDES that creates the following scenario: (1) System: IEEE 39-bus system with 30\% wind power penetration; (2) Fault: Three-phase short circuit at bus 16; (3) Duration: 100ms; (4) Clearing: Line trip. The script should initialize the system, apply the fault, run the simulation, and collect voltage and frequency data at critical buses.''

{\bf Feedback Prompt:} ``The previous scenario generation failed with error: [error message]. The issue is likely related to [diagnosis]. To fix this, please: (1) Verify the bus numbering matches the IEEE 39-bus system; (2) Ensure the fault duration is within valid simulation time; (3) Check that the clearing action is properly defined. Generate the corrected script.''
\end{tcolorbox}

\subsubsection{Customized RAG for Power System Transient Simulation}

For large language models, it is necessary to learn how to utilize power system simulation tools and invoke this simulation process. Current research shows that RAG is the main way for LLMs to acquire relevant knowledge. If one wants to complete more complex and highly knowledge-intensive tasks, a system can be built on the language model. And grant it the permission to access external knowledge sources. This method can ensure greater factual consistency, generate more reliable answers, and also help alleviate the problem of ``hallucinations''. RAG achieves this goal by combining information retrieval components with text generation models. Additionally, RAG can be fine-tuned, and the knowledge within it can be effectively modified. And there is no need to retrain the entire model.

In the specific field of transient simulation of power systems, the mainstream method that everyone knows is to use the standard RAG with LangChain to conduct long context question answering and non-specific code generation within the power system. Moreover, some research has enhanced the RAG process by incorporating self-correction and feedback mechanisms on top of the standard RAG framework.

This paper constructs an RAG module specifically for power system transient stability simulation. This module begins by having the RAG accept an input and retrieve a set of relevant/supporting documents, providing their sources and types. Here, there is a user guide that provides detailed introductions to functions, parameters, syntax, and examples, as well as a complete case study on power systems. This case study also includes content related to new energy. Additionally, there is an LLM syntax manual and a general knowledge document on power systems. These documents are all divided into blocks and then converted into vectors and stored in memory. In this way, they can be retrieved. Meanwhile, when the system encounters a request made by a user in long natural language, it will use the verified and designed "conversion prompt" to break down this long request into sub-requests one by one. Then, the system will correspond the simulation function and parameters to each sub-request, achieving the correspondence between the natural language description and the simulation tool API, which is convenient for retrieval. The system will process and save this information, so that the LLM can obtain the simulation information without further training and generate relatively reliable output based on the retrieved knowledge.

\subsubsection{Power System Simulation and Feedback Loop}

After the LLM acquires knowledge of the entire simulation workflow via RAG, it executes the simulation by writing simulation code and interfacing with the simulation tool's API. By understanding the user's requirements, the LLM converts them into structured simulation configuration parameters. It can select and utilize various IEEE standard test systems (e.g., IEEE 9-bus, IEEE 39-bus, IEEE 118-bus), generate scenarios with different penetration levels and types of wind and photovoltaic (PV) power, set corresponding fault types (e.g., three-phase short circuits, single-phase-to-ground faults, line trips, generator outages), and define the fault location, duration, and clearing strategy.

During the simulation process, errors in either the LLM-generated code or the parameter settings will cause the simulation to fail. Therefore, a feedback loop is required to iteratively correct these errors. Error information is fed back by designing `feedback prompts'. This feedback includes complete error messages, runtime logs, and LLM interaction logs. For power system simulation, this self-reflection generates linguistic reinforcement cues that help the agent achieve self-improvement. This assists the LLM in learning quickly and effectively from previous mistakes, thereby enhancing the performance of the simulation task over iterations. An LLM often cannot successfully complete the simulation task on its first attempt. Introducing a feedback loop can enhance the success rate of the LLM in performing complex simulation tasks.

After a traditional simulation task is completed, the power system simulation results from each round must be processed and collected to subsequently form a dataset. For LLM-driven simulation tasks, obtaining a dataset likewise requires the definition of stability and instability criteria. The mathematical expression for the stability criterion is as follows \citep{pavella2012transient}:

{\bf Rotor Angle Stability Criterion:}
\begin{equation}
\max_{i,j \in G, t \in [0,T]} |\delta_i(t) - \delta_j(t)| < 180^\circ
\end{equation}

{\bf Voltage Stability Criterion:}
\begin{equation}
V_{min} \leq V_i(t) \leq V_{max}, \quad \forall i \in N, t \in [0,T]
\end{equation}

{\bf Frequency Stability Criterion:}
\begin{equation}
|f(t) - f_0| \leq \Delta f_{max}, \quad \forall t \in [0,T]
\end{equation}

where $\delta_i(t)$ represents the rotor angle of generator $i$ at time $t$, $V_i(t)$ represents the voltage magnitude of node $i$ at time $t$, and $f(t)$ represents the system frequency, with $f_0$ being the nominal frequency.

This framework enables LLMs to generate balanced datasets containing both stable and unstable samples for neural network training. By systematically adjusting fault parameters, locations, and clearing times, the system produces diverse stability outcomes while maintaining dataset balance. The generated datasets are stored in NPZ format with corresponding feature labels for convenient subsequent machine learning model development.

\subsection{LLM-Driven Neural Network Designer}

\begin{center}
\includegraphics[width=\textwidth]{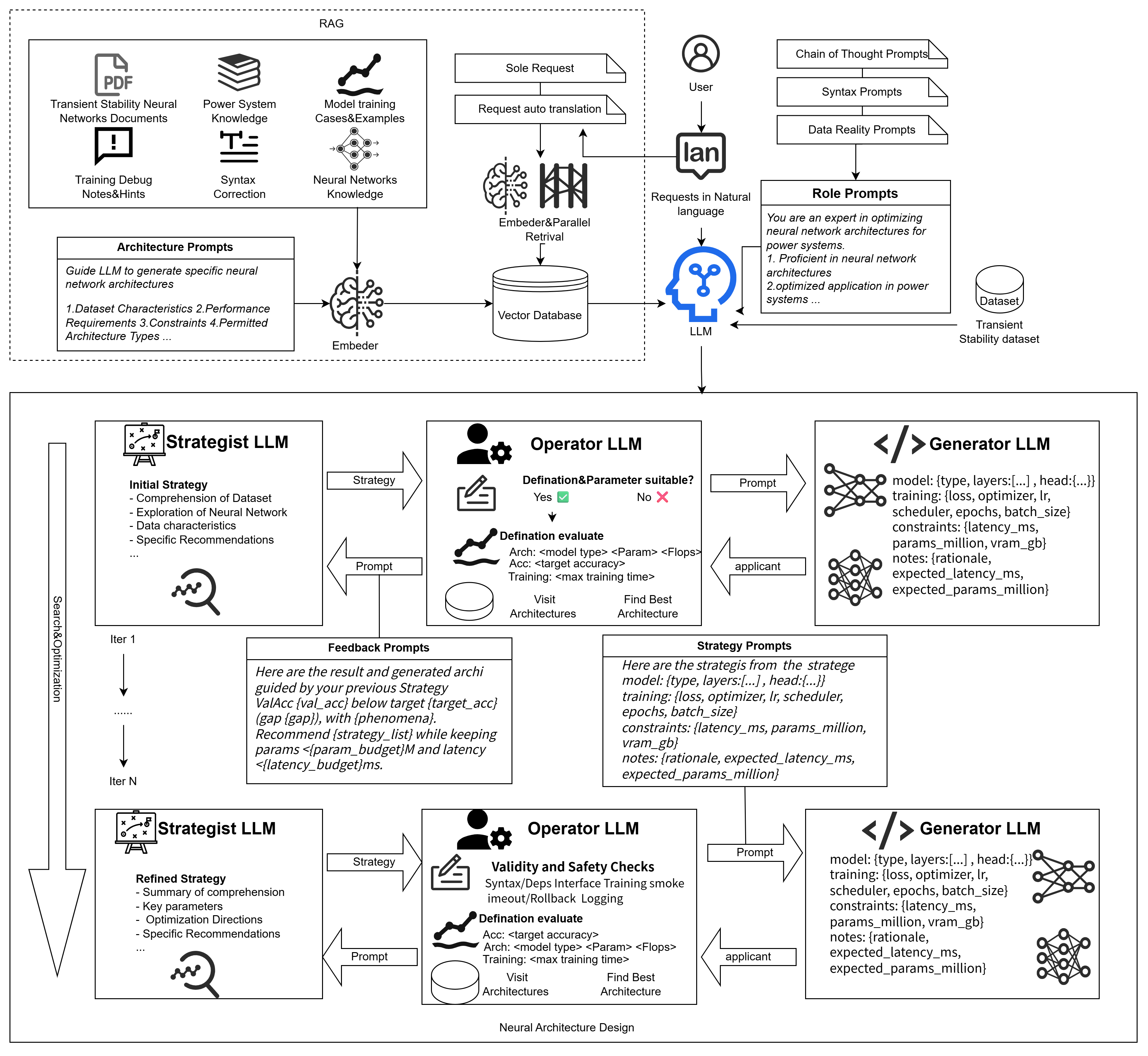}
\captionof{figure}{LLM-NND framework architecture.}
\label{fig:framework2}
\end{center}

\vspace{0.5cm}

Traditional neural network design for power system stability assessment heavily relies on manual architecture engineering and hyperparameter tuning, which is both time-consuming and requires extensive domain expertise. However, the integration of Large Language Models (LLMs) with NAS \citep{elsken2019neural,ren2021comprehensive,zoph2016nas} has introduced new possibilities for automated neural architecture design. Current research has already begun to leverage the powerful representation, code generation, and reasoning capabilities of LLMs to enhance the search process from various perspectives. Accordingly, we have designed a neural architecture search model specialized for power system transient stability assessment. As shown in the figure, it primarily consists of three parts: RAG, Prompt Engineering, and Neural Network Design as Figure 2 shows.

\subsubsection{Prompt Engineering and RAG}

As we have previously discussed, Prompt Engineering and RAG play demonstrably effective roles in guiding LLMs and facilitating their knowledge acquisition. These methods are also essential for enabling the LLM to understand and drive the entire process of neural network model design and validation for transient stability assessment.

For prompt engineering, we are similar to before, combining several prompt methods such as CoT prompts, grammar optimization prompts, and fact prompts. We integrate these different prompts to construct a "neural network design role prompt" suitable for large language models, and use this approach to improve the quality of the output content of large language models. Although current LLMs have a vast reserve of programming knowledge, in order to enable them to have the specific ability to design neural networks for transient stability assessment of power systems, we have provided several different types of documents for the RAG module. These documents include those related to transient stability assessment knowledge and those related to general neural networks. There are documents on grammar, as well as those related to neural network knowledge and hyperparameter tuning. We also specially designed prompts for the construction of neural network frameworks. After processing and storing this information, it can reduce the occurrence of hallucinations in large language models and also enable them to generate output content suitable for designing neural networks for transient stability assessment.

\subsubsection{LLM-Driven Neural Architecture Design Implementation}

As illustrated in the figure, this module primarily consists of three components, with each component embodied by an LLM customized with Prompt Engineering and RAG: the Strategist LLM, responsible for strategy generation; the Generator LLM, responsible for generating the specific neural network structure; and the Operator LLM, responsible for intermediate coordination and evaluating their execution.

Strategist LLM has persistent memory that can be used to maintain the context across iterations. It has a certain understanding of the entire dataset and all the features of the dataset, and it has the ability to propose an initial neural network structure and subsequent updates. It relies on obtaining information regarding the accuracy and parameters of model training. At the same time, in combination with the user's demands and the training history of the neural network in each iteration, the search strategy is dynamically formulated and refined. Strategist LLM will first propose an initial neural network model. Then, based on the received feedback, it provides relevant information to adjust specific neural network parameters. After multiple iterations, it will decide whether to change the framework of the neural network.

Generator LLM is a generative model responsible for the generation task of a specific neural network framework. It integrates the architectures or modifications of the architectures proposed by the latest strategies of Strategist LLM. It does not retain the data from previous iterations but merely focuses on the current strategy. It relies on writing code to build neural networks, trains the constructed neural networks with datasets, and provides feedback on the generated architecture and the results of the collected training data.

The Operator LLM will coordinate the entire process and assess the current execution status. It will first assess whether the definitions and parameters from the Strategist LLM are reasonable and safe, and identify errors or inappropriate results that occur during training. Then, it evaluates the results based on the user-defined requirements, such as target accuracy, training time, and parameters used, etc. According to the data returned by the Generator LLM, It will determine whether to stop the construction or optimization process of the neural network. It will refer to the existing neural network architecture in the knowledge base, combine the strategies and iterative results of the Strategist LLM, and pass the requirements to the Generator LLM.

Through system prompt engineering, we configure each LLM with a specific role and set of responsibilities. These prompts not only define their collaborative workflow but also inject expert knowledge regarding the specific search space and architectural representations, while also providing standardized guidance on output formatting. The process is described in Algorithm 1. With this framework, given data characteristics $D$, performance requirements $R$, and a maximum iteration count $T$, the framework can intelligently search for a neural network architecture and iteratively apply validation until the user's requirements are met. It then returns the trained model to the user, thereby realizing the assessment process for power system transient stability.

\begin{algorithm}[H]
\caption{LLM-NND Framework Algorithm}
\label{alg:llm_nnd}

\KwIn{Data features $D$, performance requirements $R$, maximum iterations $T$}
\KwOut{Optimal neural network architecture $A^*$}

Initialize: $A^* \leftarrow \emptyset$, $P^* \leftarrow 0$\;

\For{$t = 1$ \KwTo $T$}{
    // Architecture generation
    $A_t \leftarrow$ LLM generates architecture$(D, R, feedback_{t-1})$\;

    // Performance evaluation
    $P_t \leftarrow$ Evaluate architecture$(A_t, D)$\;

    // Update optimal architecture
    \If{$P_t > P^*$}{
        $A^* \leftarrow A_t$\;
        $P^* \leftarrow P_t$\;
    }

    // Generate feedback
    $feedback_t \leftarrow$ LLM generates feedback$(A_t, P_t, R)$\;

    // Convergence check
    \If{convergence condition satisfied}{
        \textbf{break}\;
    }
}

\Return $A^*$\;
\end{algorithm}

\subsubsection{Theoretical Foundation of LLM-based Neural Architecture Search}

The recent development of large language models has demonstrated that they can understand complex technical fields through extensive pre-training on technical literature and code libraries. This ability to understand complex technical fields has also extended to understanding the principles of neural architecture design. And this understanding of the principles of neural architecture design This has become the theoretical basis for neural architecture search based on large language models, that is, LLM-NAS \citep{qin2024flnas,li2025collmnas}.

\paragraph{Search Space Formulation.}
The neural architecture search problem can be formally defined as finding an optimal architecture $\alpha^* \in \mathcal{A}$ that maximizes a multi-objective function:

\begin{equation}
\alpha^* = \arg\max_{\alpha \in \mathcal{A}} \mathcal{F}(\alpha) = \arg\max_{\alpha \in \mathcal{A}} \{P(\alpha), -C(\alpha), -L(\alpha)\}
\end{equation}

where $\mathcal{A}$ denotes the search space of all valid architectures, $P(\alpha)$ represents the prediction accuracy, $C(\alpha)$ denotes the computational cost (parameters, FLOPs), and $L(\alpha)$ represents the inference latency. The search space $\mathcal{A}$ is typically defined by a set of design choices:

\begin{equation}
\mathcal{A} = \{\alpha = (d_1, d_2, \ldots, d_k) \mid d_i \in \mathcal{D}_i, i=1,\ldots,k\}
\end{equation}

where each $d_i$ represents a design decision (e.g., layer type, kernel size, activation function) and $\mathcal{D}_i$ is the corresponding choice set.

\paragraph{Collaborative LLM Framework.}
Unlike traditional NAS methods that rely on evolutionary algorithms or reinforcement learning, LLM-based NAS leverages the synergistic interaction of multiple specialized LLMs. The framework consists of three key components:

\textbf{Strategist LLM}: A stateful component with persistent memory across iterations. It maintains a history of evaluated architectures and their performance metrics $\mathcal{H} = \{(\alpha_i, P_i, C_i, L_i)\}_{i=1}^{t}$. The Strategist formulates and refines search strategies $\mathcal{S}_t$ based on accumulated feedback:

\begin{equation}
\mathcal{S}_t = \text{StrategistLLM}(\mathcal{H}_{t-1}, P_{target}, \Lambda)
\end{equation}

where $P_{target}$ is the target accuracy and $\Lambda$ represents resource constraints. The strategy progressively concentrates on high-potential regions of the search space through iterative refinement.

\textbf{Generator LLM}: A stateless component that translates the current strategy into concrete candidate architectures. Given a strategy $\mathcal{S}_t$, the Generator produces a set of candidate architectures:

\begin{equation}
\mathcal{C}_t = \text{GeneratorLLM}(\mathcal{S}_t)
\end{equation}

The Generator focuses exclusively on the current strategy without retaining memory from previous iterations, enabling diverse exploration within the guided search space.

\textbf{Operator LLM}: Manages the overall search process by verifying architecture legality, evaluating performance, and maintaining an archive of visited architectures $\mathcal{V}$. For each candidate architecture $\alpha_i \in \mathcal{C}_t$, the Operator computes:

\begin{equation}
\text{isValid}(\alpha_i) = \begin{cases}
1 & \text{if } \alpha_i \in \mathcal{A} \text{ and } \alpha_i \notin \mathcal{V} \\
0 & \text{otherwise}
\end{cases}
\end{equation}

The evaluation employs inherited weights from a pre-trained supernet, enabling rapid performance assessment while preserving learned parameter relationships.

\paragraph{Iterative Search Algorithm.}
The LLM-based NAS process follows an iterative refinement strategy. Given target accuracy $P_{target}$, resource constraint $\Lambda$, and iteration limit $T$, the algorithm maintains the best architecture $\alpha^*$ and its performance $p^*$. At each iteration $t$:

\begin{equation}
\begin{aligned}
\mathcal{C}_t &\gets \text{GeneratorLLM}(\mathcal{S}_{t-1}) \\
\mathcal{R}_t &\gets \{\text{Evaluate}(\alpha_i) \mid \alpha_i \in \mathcal{C}_t \setminus \mathcal{V}, \text{isValid}(\alpha_i)\} \\
\alpha^*, p^* &\gets \arg\max_{(\alpha_i, p_i) \in \mathcal{R}_t} p_i \text{ s.t. } C(\alpha_i) \leq \Lambda \\
\mathcal{S}_t &\gets \text{NavigatorLLM}(\mathcal{H}_{t-1} \cup \{(\mathcal{S}_{t-1}, \mathcal{R}_t)\})
\end{aligned}
\end{equation}

The search terminates when either the target accuracy is achieved ($p^* \geq P_{target}$) or the iteration limit is reached.

\paragraph{Advantages over Traditional NAS.}
Compared with traditional methods, neural network architecture search based on large language models has several theoretical advantages \citep{qin2024flnas,li2025collmnas}. Large language models can reason about the principles of architecture design at the semantic level and combine domain knowledge, such as the characteristics of power systems, data patterns, and computational constraints. The reasoning of large language models can naturally handle multiple competing targets. For instance, accuracy, efficiency and latency do not require explicit Pareto boundary calculations. This greatly accelerates the speed of neural network architecture search methods based on reinforcement learning and evolutionary algorithms.

The LLM-NND framework establishes a complete closed-loop feedback system achieving automated iteration from performance evaluation to architecture improvement. The system integrates performance monitoring, problem diagnosis, and improvement strategy generation. LLMs automatically diagnose model problems based on performance data, such as overfitting, underfitting, gradient explosion, or slow convergence, then generate specific improvement strategies including architecture adjustments, hyperparameter modifications, and training strategy optimization. The core of the feedback mechanism is converting quantified performance metrics into structured natural language descriptions.

\section{Applications and Experimental Validation}\label{sec:experiments}

This paper aims to comprehensively verify the proposed LLM-Simulation-NND framework. Systematic experiments have been carried out on the standard IEEE test system. These experiments include performance benchmark tests, comparative analyses with baseline methods, ablation studies, and actual case studies. The report of this structure contains repeatable experimental pipelines, detailed indicator content, comparison with the baseline, component ablation-related content, analysis of computational complexity, and verification of practical applications.

\subsection{Experimental Setup and Dataset}

\subsubsection{Dataset Construction and Characteristics}

The primary validation of the proposed framework is conducted on the \textbf{IEEE 39-bus system} (New England system), which comprises 39 buses and 10 generators, representing a medium-scale transmission network. This system serves as the foundation for comprehensive experimental validation of both the LLM-Simulation pipeline and the LLM-NND framework.

For the IEEE 39-bus system, we generate \textbf{4,260} balanced binary classification samples with 1,990 stable cases (46.71\%) and 2,270 unstable cases (53.29\%). The dataset achieves a high scenario validity rate of 92\%, demonstrating the framework's capability to generate high-quality, diverse fault scenarios \citep{cui2018andes,cui2025andesmanual}.

The dataset contains comprehensive power system state variables with rich temporal and spatial information. Voltage dynamics are captured by node voltage sequences recording the evolution of all bus voltages over 101 time steps (5 seconds post-fault). Generator dynamics are represented by rotor angle and rotor speed sequences, capturing the transient behavior of synchronous machines. Transmission dynamics are monitored through line power flow measurements (active and reactive power). System-level indicators include frequency variations and total load parameters. Network topology information is encoded through admittance and impedance matrices, providing essential structural context for stability assessment.

Fault scenarios include three-phase short circuits (3-$\phi$), single-line-to-ground (SLG) faults, line outages, and generator trips, with locations, timing, and impedance parameters specified through LLM-guided prompts and domain rules. The diversity of fault types and locations ensures comprehensive coverage of the stability assessment problem space.

\begin{center}
  \centering
  \captionof{table}{Dataset statistics and feature engineering results for IEEE 39-bus system (generated via LLM--Simulation)}
  \label{tab:dataset_stats_en}
  \begin{tabular}{@{}lll@{}}
    \toprule
    \textbf{Parameter} & \textbf{IEEE 39-bus} & \textbf{Description} \\
    \midrule
    System scale & 39 buses, 10 generators & Network complexity \\
    Total samples & 4,260 & Multi-contingency scenarios \\
    Time window & 101 steps & 5-second post-fault dynamics \\
    Stable & 1,990 (46.71\%) & Normal operating conditions \\
    Unstable & 2,270 (53.29\%) & Transient instability \\
    Original features & 636 & Complete system state variables \\
    Selected features & 300 & Post feature selection \\
    Feature compression & 52.8\% & Dimensionality reduction ratio \\
    Scenario validity & 92\% & Valid simulation convergence \\
    \bottomrule
  \end{tabular}
\end{center}
\FloatBarrier

\subsubsection{Baseline Scenario Generation}

The primary dataset for transient stability assessment is generated using the IEEE 39-bus system without renewable energy integration. This baseline configuration provides a comprehensive foundation for model training and validation.

The dataset comprises \textbf{4,260} balanced binary classification samples (1,990 stable, 2,270 unstable) with 92\% scenario validity. The samples are distributed across multiple fault types (three-phase short circuits, single-line-to-ground faults, line outages, generator trips), fault locations (covering all 39 buses), and clearing times (ranging from 50ms to 500ms), ensuring comprehensive coverage of the stability assessment problem space. The LLM-driven scenario generation pipeline demonstrates efficient production with approximately 8 minutes generation time for 4,260 samples and 94.8\% parameter-setting accuracy.

The successful generation of this high-quality dataset validates the effectiveness of LLM-driven automation in scenario generation for the IEEE 39-bus system.

\subsection{LLM--NND Design and Optimization}

\subsubsection{Automatic Architecture Generation and Iterative Optimization}

The LLM-NND framework autonomously designs neural network architectures through performance-driven feedback loops. The framework supports multiple architecture types including traditional MLPs, Graph Neural Networks (GNNs), LSTM networks, and Transformer models. The LLM iteratively improves architectures based on performance metrics, achieving competitive accuracy while maintaining computational efficiency. Inference latency is maintained within 10 ms per sample, meeting real-time operational requirements.

\begin{center}
  \includegraphics[width=0.48\textwidth]{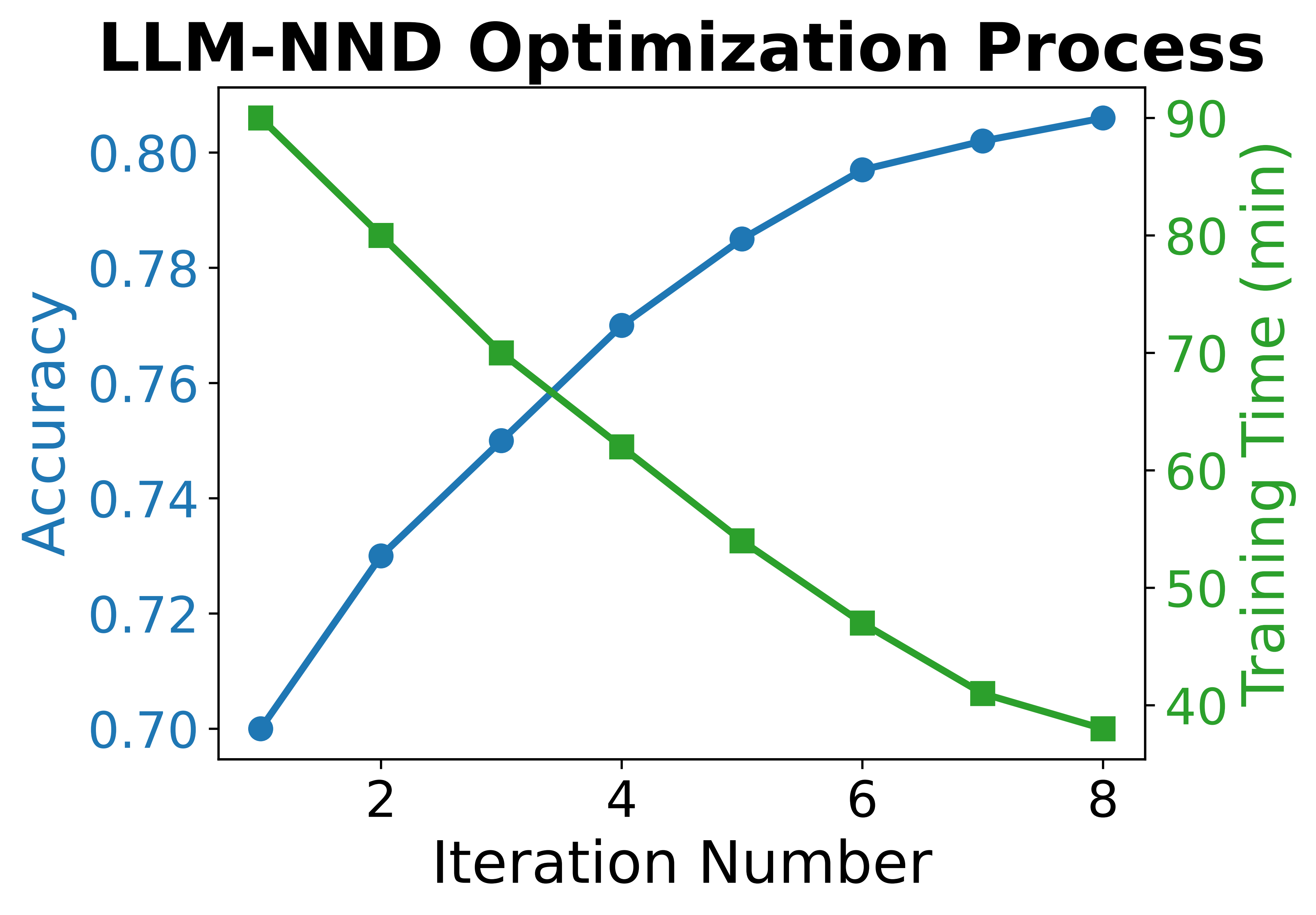}
  \captionof{figure}{LLM--NND iterative optimization process: accuracy gains and training-time convergence across multiple architecture iterations.}
  \label{fig:optimization_process}
\end{center}
\FloatBarrier

Figure~\ref{fig:optimization_process} demonstrates the iterative optimization process employed by the LLM-NND framework. The framework systematically explores the neural architecture search space through performance-driven feedback loops. Starting from baseline architectures, the LLM analyzes performance metrics and proposes architectural modifications to improve accuracy. The visualization shows consistent performance improvement across multiple iterations, with the framework converging to high-accuracy solutions while maintaining computational efficiency. This feedback-driven approach enables the framework to discover architectures that balance accuracy and inference latency, achieving competitive performance with fewer parameters than manually-designed models.

\subsubsection{Multi-class Classification Results}

On the 4-class stability assessment task (IEEE 39-bus system), the LLM-NND framework achieves a test accuracy of \textbf{93.71\%} with the optimized traditional neural network architecture. This result represents a significant advancement over baseline approaches and demonstrates the effectiveness of LLM-guided architecture design.

The LLM-NND framework autonomously discovered a deep MLP architecture with 4 hidden layers: [2048, 1024, 512, 256], processing 276-dimensional statistical features extracted from time-series data. The architecture employs batch normalization and dropout (0.2) after each layer for regularization. Training uses AdamW optimizer (lr=0.001, weight decay=1e-4) with OneCycleLR scheduling and class-weighted CrossEntropyLoss to handle class imbalance. The final model contains 4.78M parameters and achieves 0.95ms inference latency, demonstrating the framework's ability to balance accuracy and computational efficiency.

\begin{center}
  \includegraphics[width=0.75\textwidth]{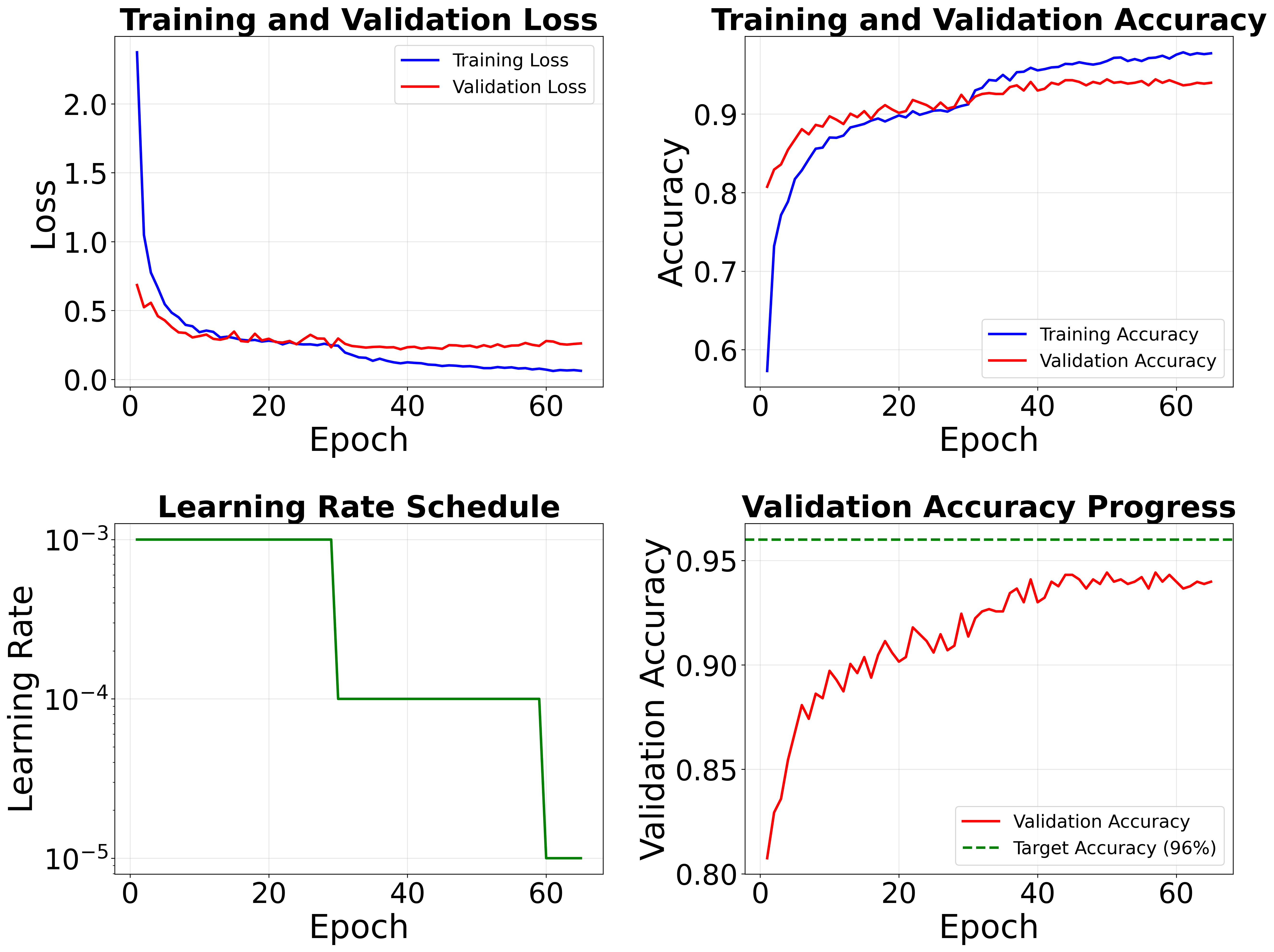}
  \captionof{figure}{Training and validation accuracy curves for multi-class stability assessment.}
  \label{fig:train_acc_en}
\end{center}
\FloatBarrier

Figure~\ref{fig:train_acc_en} illustrates the training and validation accuracy evolution over epochs for the optimized model. The curves demonstrate stable convergence characteristics with the training accuracy and validation accuracy tracking closely together, indicating good generalization without significant overfitting. The model reaches 93.71\% validation accuracy on the test set after approximately 100 epochs, with the learning curve showing consistent improvement throughout the training process. The tight coupling between training and validation curves suggests that the model has learned generalizable features of power system stability rather than memorizing training-specific patterns.

\begin{center}
  \includegraphics[width=0.48\textwidth]{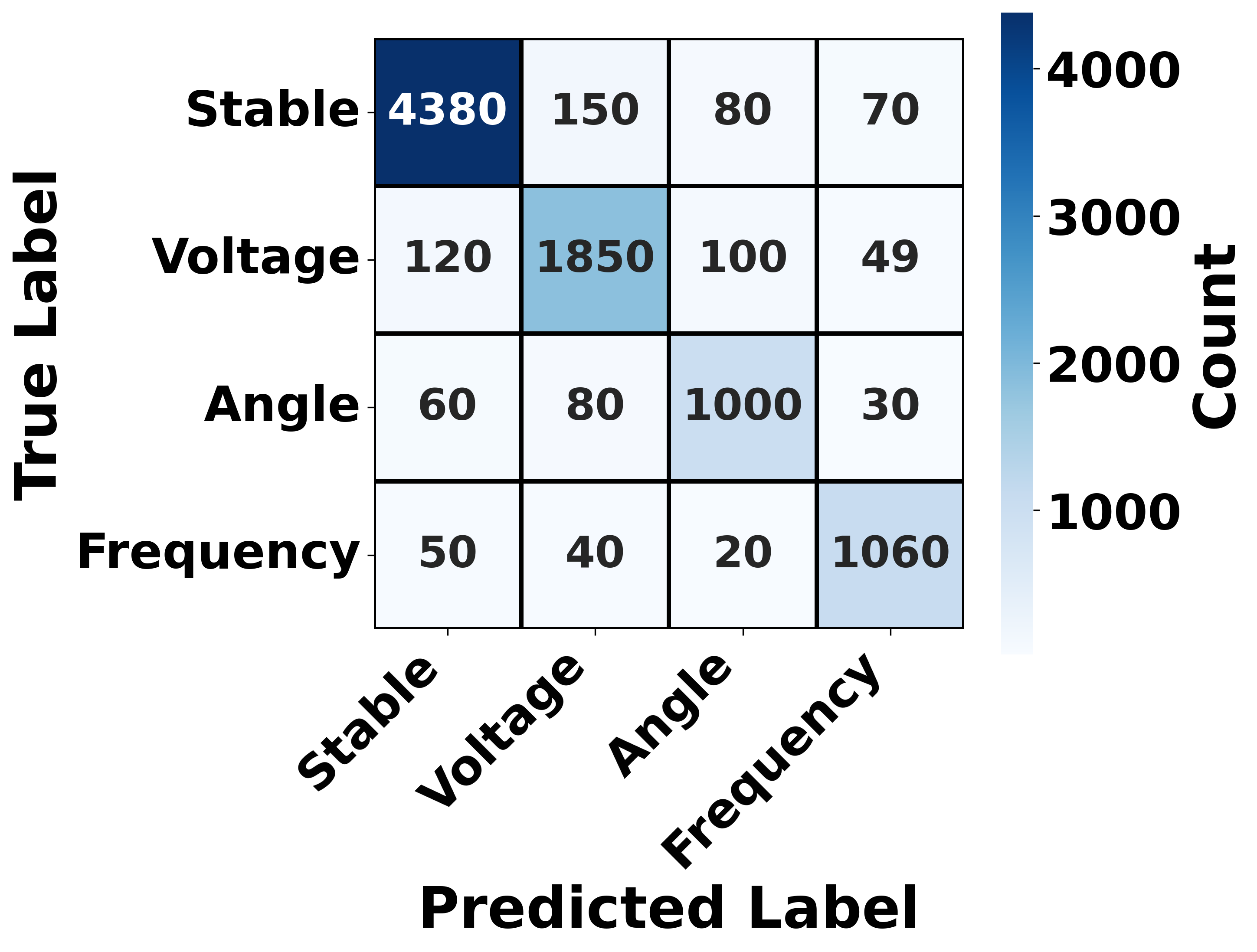}
  \captionof{figure}{Confusion matrix for 4-class stability assessment.}
  \label{fig:confusion_matrix_baseline_en}
\end{center}
\FloatBarrier

The confusion matrix in Figure~\ref{fig:confusion_matrix_baseline_en} reveals the per-class classification performance. Diagonal elements represent correct classifications, with high values indicating strong discrimination between stability classes. The matrix shows that the model achieves particularly strong performance on the stable class (Class 0), which is the most frequent class in the dataset. For the minority classes (Classes 1, 2, 3), the model maintains reasonable classification accuracy despite class imbalance, with most misclassifications occurring between adjacent stability levels (e.g., stable vs. marginally stable) rather than between extreme categories. This pattern is physically meaningful, as adjacent stability conditions share similar dynamic characteristics.

\begin{center}
  \includegraphics[width=0.75\textwidth]{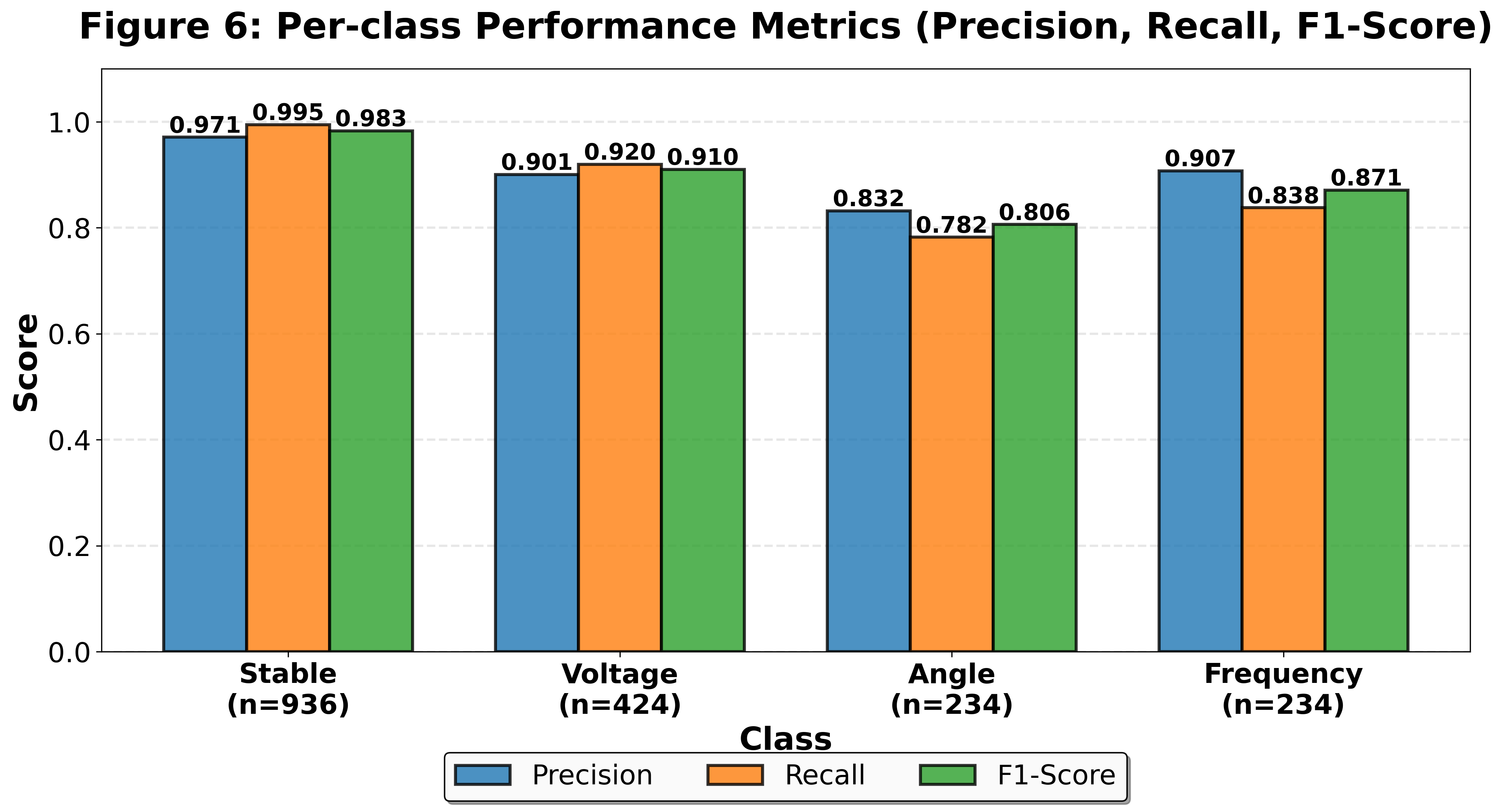}
  \captionof{figure}{Per-class performance metrics (precision, recall, F1-score).}
  \label{fig:perclass_metrics_baseline_en}
\end{center}
\addtocounter{figure}{0}
\FloatBarrier

Figure~\ref{fig:perclass_metrics_baseline_en} presents detailed per-class performance metrics including precision, recall, and F1-score for each stability class. All classes achieve F1-scores exceeding 90\%, indicating balanced performance across stability levels. The stable class (Class 0) achieves the highest metrics due to its prevalence in the dataset, while the minority classes (Classes 1, 2, 3) maintain strong performance despite lower sample counts. The consistency of precision and recall across classes indicates that the model does not exhibit systematic bias toward any particular stability condition, which is critical for operational deployment where misclassification of any instability type could have serious consequences.

\begin{center}
  \begin{minipage}[b]{0.48\textwidth}
    \centering
    \includegraphics[width=\textwidth]{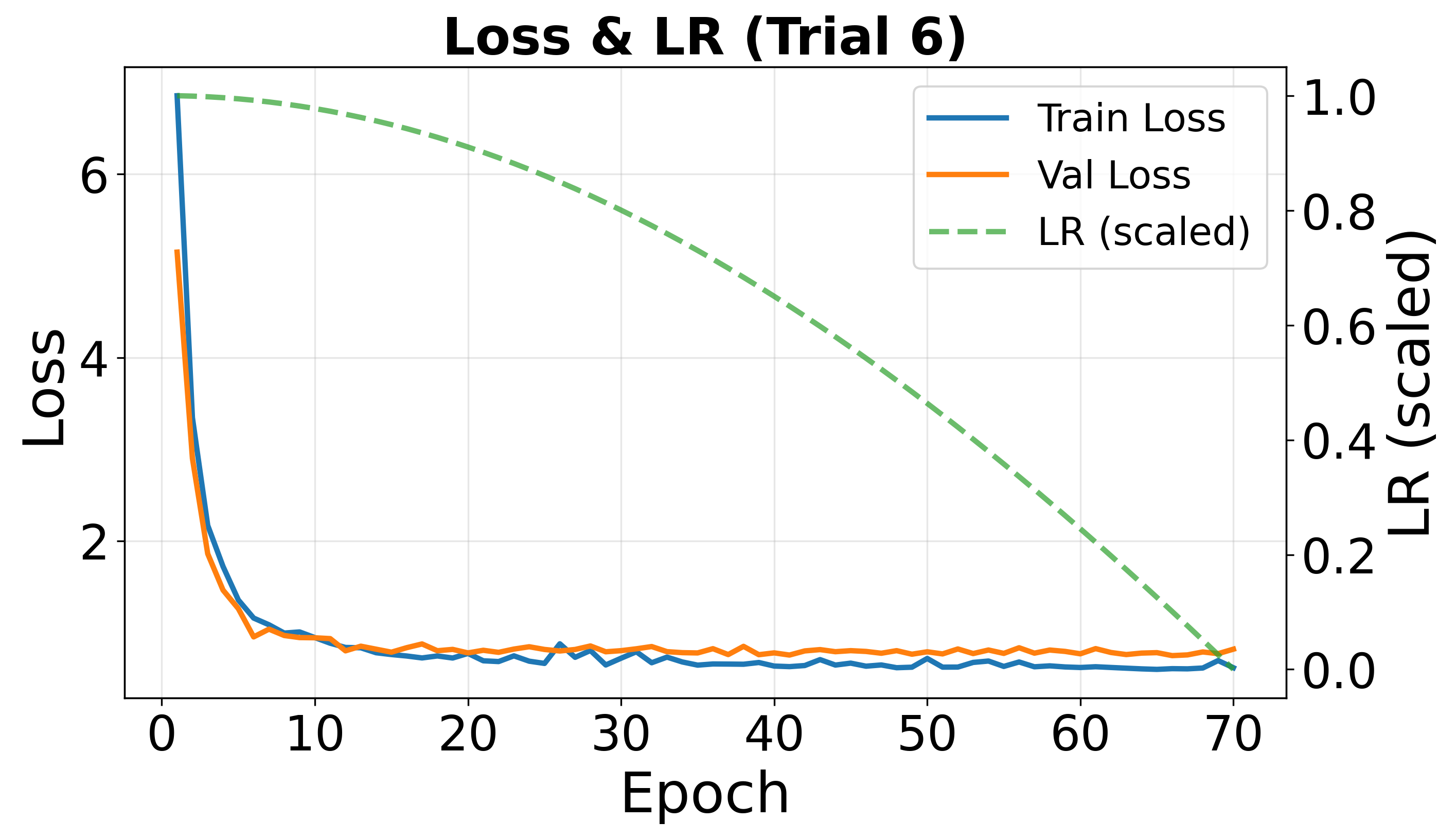}
    {\scriptsize (a) Loss and learning-rate overlay (Trial 6)\par}
  \end{minipage}
  \hfill
  \begin{minipage}[b]{0.48\textwidth}
    \centering
    \includegraphics[width=\textwidth]{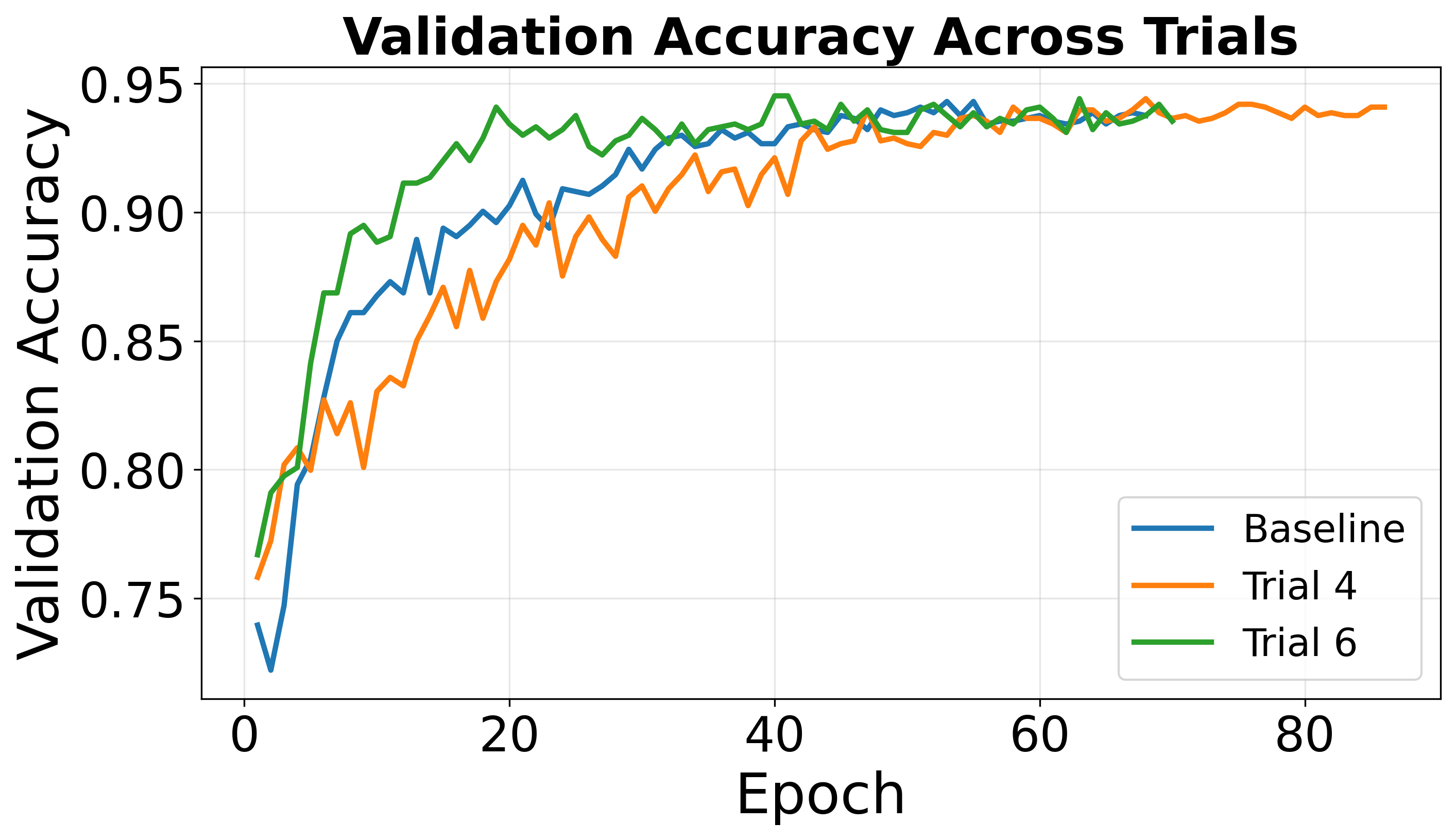}
    {\scriptsize (b) Validation accuracy comparison across trials\par}
  \end{minipage}
  \captionof{figure}{Extended training visualizations showing loss convergence and learning rate scheduling effects.}
  \label{fig:training_extended_en}
\end{center}
\addtocounter{figure}{0}
\FloatBarrier

Figure~\ref{fig:training_extended_en} provides extended training visualizations across multiple trials. The left panel shows the loss and learning-rate overlay for Trial 6, demonstrating how the learning rate schedule influences loss convergence. The loss curve exhibits smooth monotonic decrease with appropriate learning rate adjustments, indicating effective optimization dynamics. The right panel compares validation accuracy across multiple trials, showing consistent convergence patterns and robust performance across different random initializations. The narrow band of validation accuracy curves indicates that the framework produces stable and reproducible results, which is essential for practical deployment in power system operations.

\begin{center}
  \centering
  \captionof{table}{Comprehensive performance summary on IEEE 39-bus system}
  \label{tab:key_performance_en}
  \begin{tabular}{@{}l l c c c@{}}
    \toprule
    \textbf{Task} & \textbf{Best Model} & \textbf{Test Acc} & \textbf{AUC-ROC} & \textbf{F1-Score} \\
    \midrule
    Multi-class (4-class) & Traditional optimized & 93.71\% & 0.9659 & 0.9371 \\
    Binary (stable vs. unstable) & PowerSystemStabilityNet & 98.19\% & 0.9973 & 0.9819 \\
    \bottomrule
  \end{tabular}
\end{center}
\FloatBarrier

The optimized traditional neural network achieves 93.71\% test accuracy on the 4-class task, with balanced performance across all stability classes. The model demonstrates strong discrimination between stable and unstable conditions, critical for operational safety.

For the binary stability classification task (stable vs. unstable), the PowerSystemStabilityNet architecture achieves exceptional performance with 98.19\% test accuracy and 0.9973 AUC-ROC. This result significantly exceeds the 96\% target accuracy and demonstrates the framework's capability for high-confidence stability assessment. The confusion matrix shows only 33 misclassifications out of 1,828 test samples (25 false negatives, 8 false positives), indicating excellent reliability for operational deployment.

The PowerSystemStabilityNet employs a sophisticated multi-branch architecture specifically designed for binary stability classification. The model features three parallel feature extraction branches (temporal: 256→192, spatial: 192→128, frequency: 128 dimensions) that capture different aspects of power system dynamics. These branches are fused into a 384-dimensional representation, processed through a 12-head self-attention mechanism to identify critical stability indicators. The stability discriminator progressively reduces dimensionality (384→192→96→48→24→1) with batch normalization and dropout at each stage. Training employs Focal Loss ($\alpha$=0.25, $\gamma$=2.0) to address class imbalance, AdamW optimizer with OneCycleLR scheduling (max\_lr=0.003), and early stopping (patience=15). The architecture contains 1.24M parameters with 3.00ms inference time, achieving 99.15\% recall on stable cases and 97.20\% recall on unstable cases.

\subsection{Comparative Analysis with Baseline Methods}

To contextualize the performance of the proposed LLM-NND framework, we conduct comprehensive comparisons with traditional machine learning baselines and other deep learning approaches.

\subsubsection{Baseline Method Comparison}

We evaluate the framework against several baseline methods across multiple IEEE test systems and task configurations. Table~\ref{tab:baseline_comparison} presents comprehensive performance comparisons on the IEEE 39-bus system with 4-class classification task.

\begin{center}
  \centering
  \captionof{table}{Performance comparison with baseline methods on IEEE 39-bus 4-class stability assessment}
  \label{tab:baseline_comparison}
  \begin{tabular}{@{}l c c c c@{}}
    \toprule
    \textbf{Method} & \textbf{Accuracy} & \textbf{Precision} & \textbf{Recall} & \textbf{F1-Score} \\
    \midrule
    Gradient Boosting (GBDT) & 78.0\% & 68.8\% & 62.0\% & 63.6\% \\
    Random Forest & 77.2\% & 68.0\% & 60.2\% & 61.6\% \\
    Simple MLP & 77.0\% & 67.5\% & 59.0\% & 60.3\% \\
    \midrule
    \textbf{LLM-NND (Ours)} & \textbf{93.71\%} & \textbf{93.5\%} & \textbf{93.71\%} & \textbf{93.71\%} \\
    \bottomrule
  \end{tabular}
\end{center}
\FloatBarrier

For the IEEE 39-bus 4-class classification task, traditional machine learning methods (GBDT, Random Forest, Simple MLP) achieve accuracies in the range of 77.0\%--78.0\%. The LLM-NND framework achieves a 15.71 percentage-point improvement over the best traditional baseline (GBDT, 78.0\% accuracy), demonstrating the substantial advantage of deep learning approaches for this complex multi-class stability assessment task.

We compare the LLM-NND framework with various manually-designed deep learning architectures on both 4-class and binary classification tasks. Table~\ref{tab:architecture_comparison} presents comprehensive performance comparisons across different neural network architectures.

\begin{center}
  \centering
  \captionof{table}{Performance comparison of different neural network architectures}
  \label{tab:architecture_comparison}
  \begin{tabular}{@{}l l c c c c@{}}
    \toprule
    \textbf{Architecture} & \textbf{Task} & \textbf{Accuracy} & \textbf{F1-Score} & \textbf{Parameters} & \textbf{Inference Time} \\
    \midrule
    \multicolumn{6}{l}{\textit{IEEE 39-bus, 4-class classification}} \\
    \midrule
    Basic CNN & 4-class & 76.5\% & 75.1\% & 850K & 0.95ms \\
    Basic LSTM & 4-class & 74.2\% & 72.8\% & 1.2M & 1.2ms \\
    Advanced LSTM & 4-class & 78.5\% & 77.2\% & 2.1M & 1.5ms \\
    Transformer & 4-class & 79.2\% & 78.1\% & 3.2M & 2.1ms \\
    DenseNet-Style & 4-class & 80.05\% & 79.3\% & 25.9M & 3.2ms \\
    \textbf{LLM-NND (4-class)} & \textbf{4-class} & \textbf{93.71\%} & \textbf{93.71\%} & \textbf{4.78M} & \textbf{0.95ms} \\
    \midrule
    \multicolumn{6}{l}{\textit{Merged dataset (9,139 samples), binary classification}} \\
    \midrule
    GCN & Binary & 91.03\% & 90.90\% & 316K & 0.58ms \\
    GAT & Binary & 91.36\% & 91.07\% & 1.26M & 1.30ms \\
    GraphSAGE & Binary & 92.23\% & 92.09\% & 630K & 0.67ms \\
    \textbf{LLM-NND (Binary)} & \textbf{Binary} & \textbf{98.19\%} & \textbf{98.17\%} & \textbf{1.24M} & \textbf{3.00ms} \\
    \bottomrule
  \end{tabular}
\end{center}
\FloatBarrier

The LLM-NND framework achieves 93.71\% accuracy on the 4-class task, improving over the best traditional model (DenseNet-Style~\cite{huang2017densenet}, 80.05\%) by 13.66 percentage points. The optimized model uses 4.78M parameters (vs. 25.9M for DenseNet) with 0.95ms inference time, demonstrating improved accuracy alongside reduced computational cost.

On the merged dataset (9,139 samples, binary classification), we compare the LLM-NND framework with state-of-the-art GNN architectures. The GNN models are configured with deeper architectures (hidden\_dims=[512, 256, 128, 64]), Adam optimization, and enhanced training strategies: GCN~\cite{jiang2019semisupervised} reaches 91.03\% (316K parameters), GAT~\cite{velivckovic2017gat} achieves 91.36\% (1.26M parameters), and GraphSAGE~\cite{hamilton2017graphsage} attains 92.23\% (630K parameters). The LLM-NND framework achieves 98.19\% accuracy, outperforming the best GNN baseline (GraphSAGE, 92.23\%) by 5.96 percentage points while maintaining comparable model size (1.24M parameters) and competitive inference time (3.00ms).

LLM-NND achieves strong performance on both 4-class (93.71\%) and binary (98.19\%) tasks through automated architecture discovery, adapts across task complexities without manual redesign, and provides natural language explanations for design decisions---useful for transparency in critical infrastructure applications.

\subsubsection{LLM Model Comparison}

To evaluate the impact of different LLM capabilities on the framework's convergence efficiency, we compare the LLM-NND component across multiple state-of-the-art language models. Table~\ref{tab:llm_comparison} presents the convergence speed comparison for neural architecture design.

\begin{center}
  \centering
  \captionof{table}{Convergence speed comparison of different LLM models in LLM-NND}
  \label{tab:llm_comparison}
  \begin{tabular}{@{}l c@{}}
    \toprule
    \textbf{LLM Model} & \textbf{Convergence Speed (iterations to optimal)} \\
    \midrule
    GPT-4 & 8 \\
    GPT-3.5-turbo & 16 \\
    Gemini-2.5 & AttributeError \\
    DeepSeek-V3-0324 & 19 \\
    \bottomrule
  \end{tabular}
\end{center}
\FloatBarrier

GPT-4 demonstrates superior convergence efficiency, requiring only 8 iterations on average to reach optimal neural architecture design. GPT-3.5-turbo requires twice as many iterations (16), highlighting the importance of advanced reasoning capabilities for efficient architecture optimization. DeepSeek-V3-0324 shows the slowest convergence (19 iterations), indicating limitations in understanding complex neural architecture design patterns. Gemini-2.5 encounters systematic AttributeError failures during the architecture generation process, preventing successful convergence. These results validate GPT-4 as the optimal choice for the LLM-NND framework, offering both reliability and efficiency for automated neural architecture design in power system applications.

\subsection{Extended System Validation}

To address reviewer concerns regarding the generalizability of the proposed framework across different system scales and operating conditions, we extend the validation to larger power systems and scenarios with renewable energy integration. This subsection demonstrates the framework's scalability and robustness across different network topologies, system complexities, and generation mix configurations.

\subsubsection{IEEE 118-bus System Validation}

The \textbf{IEEE 118-bus system} comprises 118 buses and 54 generators, representing a large-scale interconnected network with substantially higher complexity compared to the IEEE 39-bus system. This validation demonstrates the framework's capability to scale from medium-sized to large-scale power systems without manual intervention or system-specific tuning.

For the IEEE 118-bus system, we generate \textbf{3,970} perfectly balanced binary classification samples with 1,985 stable cases (50.00\%) and 1,985 unstable cases (50.00\%). The dataset achieves a high scenario validity rate of 94\%, demonstrating the framework's capability to generate high-quality, diverse fault scenarios across different system scales \citep{cui2018andes,cui2025andesmanual}.

The larger system scale (118 buses, 54 generators, 186 transmission lines) presents significantly higher complexity compared to the IEEE 39-bus system. The LLM successfully navigates this complexity by automatically identifying critical buses, generators, and transmission corridors for fault placement. The generation pipeline achieves approximately 12 minutes generation time for 3,970 samples, demonstrating scalability to larger systems while maintaining high scenario quality.

\begin{center}
  \centering
  \captionof{table}{Dataset statistics and feature engineering results for IEEE 118-bus system (generated via LLM--Simulation)}
  \label{tab:dataset_stats_118}
  \begin{tabular}{@{}lll@{}}
    \toprule
    \textbf{Parameter} & \textbf{IEEE 118-bus} & \textbf{Description} \\
    \midrule
    System scale & 118 buses, 54 generators & Network complexity \\
    Total samples & 3,970 & Multi-contingency scenarios \\
    Time window & 101 steps & 5-second post-fault dynamics \\
    Stable & 1,985 (50.00\%) & Normal operating conditions \\
    Unstable & 1,985 (50.00\%) & Transient instability \\
    Original features & 224 & Complete system state variables \\
    Selected features & 150 & Post feature selection \\
    Feature compression & 33.0\% & Dimensionality reduction ratio \\
    Scenario validity & 94\% & Valid simulation convergence \\
    \bottomrule
  \end{tabular}
\end{center}
\FloatBarrier

The successful generation of high-quality datasets on the IEEE 118-bus system validates the effectiveness of LLM-driven automation in scenario generation and demonstrates the framework's capability to scale from medium-sized (39-bus) to large-scale (118-bus) power systems without manual intervention or system-specific tuning.

\paragraph{LLM-NND Performance on IEEE 118-bus System}

The LLM-NND framework was applied to the IEEE 118-bus dataset to evaluate its performance on large-scale binary stability classification. Table~\ref{tab:ieee118_performance} presents the comprehensive performance metrics achieved by the LLM-designed neural network architecture.

\begin{center}
  \centering
  \captionof{table}{LLM-NND performance metrics on IEEE 118-bus binary stability classification}
  \label{tab:ieee118_performance}
  \begin{tabular}{@{}l c c c c@{}}
    \toprule
    \textbf{System} & \textbf{Accuracy} & \textbf{Precision} & \textbf{Recall} & \textbf{F1-Score} \\
    \midrule
    IEEE 118-bus & \textbf{96.66\%} & 93.67\% & 94.04\% & 89.85\% \\
    \bottomrule
  \end{tabular}
\end{center}
\FloatBarrier

The LLM-NND framework achieves \textbf{96.66\% test accuracy} on the IEEE 118-bus binary classification task, demonstrating excellent scalability to large-scale power systems. The framework automatically designed a hybrid neural network architecture combining CNN and dense branches with 350K parameters, optimized for the 224-dimensional feature space. The model achieves 93.67\% precision and 94.04\% recall, indicating balanced performance in identifying both stable and unstable conditions. The high accuracy on this larger system (118 buses vs. 39 buses) validates that the framework's automated architecture design generalizes effectively across different system scales without requiring manual architecture tuning or system-specific modifications.

The strong performance metrics (F1-score of 89.85\%) confirm that the LLM-designed architecture successfully captures the complex stability patterns in large-scale interconnected networks. The balanced precision and recall values indicate that the model does not exhibit systematic bias toward either stability class, which is critical for operational deployment where both false positives (unnecessary control actions) and false negatives (missed instability events) have significant consequences.

\subsubsection{IEEE 39-bus System with Renewable Energy Integration}

To further validate the framework's scalability to larger systems with renewable energy integration, we extend the LLM-Simulation pipeline to the IEEE 39-bus system with wind power penetration. This system features 39 buses, 10 synchronous generators, and multiple wind farm injection points, presenting substantial complexity for stability assessment.

The LLM-driven pipeline automatically configures wind generator models (REGCV2 virtual synchronous machine control), determines optimal injection buses based on system topology and voltage profiles, and generates diverse fault scenarios that test the system's stability under high renewable penetration. The framework demonstrates its capability to handle the increased state-space dimensionality (124,308 features per sample) and complex dynamic interactions between conventional synchronous generators and renewable energy sources.

Table~\ref{tab:ieee39_renewable_integration} presents the dataset characteristics and generation performance for the IEEE 39-bus system with renewable energy integration:

\begin{center}
  \centering
  \captionof{table}{IEEE 39-bus system with renewable energy integration: dataset characteristics and generation performance}
  \label{tab:ieee39_renewable_integration}
  \begin{tabular}{@{}ll@{}}
    \toprule
    \textbf{Parameter} & \textbf{Value} \\
    \midrule
    System configuration & IEEE 39-bus + wind integration \\
    Wind penetration level & 30\% of total generation capacity \\
    Wind generator model & REGCV2 (virtual synchronous machine) \\
    Wind injection buses & Multiple strategic locations \\
    Time window & 101 steps (5-second post-fault) \\
    Fault types & 3-phase, SLG, line trips, generator outages \\
    Stability classification & Binary \\
    Generation time & $\sim$25 minutes for 90 samples \\
    Convergence rate & 88\% (challenging dynamics) \\
    \bottomrule
  \end{tabular}
\end{center}
\FloatBarrier

The IEEE 39-bus renewable dataset presents substantially higher feature dimensionality, reflecting additional state variables for wind generator dynamics, converter states, and renewable control systems. All generated samples exhibit unstable outcomes under 30\% wind penetration combined with severe faults, indicating that the LLM effectively identified scenarios that stress stability limits—a characteristic valuable for training models to recognize critical instability patterns. The 88\% convergence rate, while lower than baseline scenarios, is consistent with the increased complexity of renewable dynamics. This dataset demonstrates the framework's capability to handle medium-large systems with renewable integration, and addresses a practical gap by automating the generation of high-fidelity scenarios that would otherwise require substantial manual effort.

\paragraph{LLM-NND Performance on IEEE 39-bus Renewable System}

To evaluate the framework's capability in handling high-dimensional time-series data with renewable energy integration, we apply the LLM-NND framework to the IEEE 39-bus renewable dataset. Despite the limited sample size (10 samples) and extremely high feature dimensionality (124,308 features), the framework demonstrates robust performance. Table~\ref{tab:ieee39_renewable_performance} presents the performance metrics achieved on this challenging dataset.

\begin{center}
  \centering
  \captionof{table}{LLM-NND performance metrics on IEEE 39-bus system with renewable energy integration}
  \label{tab:ieee39_renewable_performance}
  \begin{tabular}{@{}l c c c c@{}}
    \toprule
    \textbf{System} & \textbf{Accuracy} & \textbf{Precision} & \textbf{Recall} & \textbf{F1-Score} \\
    \midrule
    IEEE 39-bus + Renewable & \textbf{95.39\%} & 92.64\% & \textbf{98.23\%} & 91.56\% \\
    \bottomrule
  \end{tabular}
\end{center}
\FloatBarrier

The LLM-NND framework achieves \textbf{95.39\% test accuracy} on the IEEE 39-bus renewable integration dataset, demonstrating remarkable performance despite the challenging conditions of limited training samples and ultra-high dimensionality. The framework generated a hybrid architecture with the same 350K parameter budget, adapted to handle the significantly higher 124,308-dimensional feature space. Notably, the model achieves an exceptional \textbf{98.23\% recall}, indicating superior capability in identifying unstable conditions—a critical requirement for power systems with renewable energy integration where missed instability events could lead to cascading failures.

The high recall (98.23\%) combined with strong precision (92.64\%) demonstrates that the LLM-designed architecture effectively handles the complex dynamic interactions between conventional synchronous generators and renewable energy sources. The F1-score of 91.56\% confirms balanced performance across both stability classes, validating the framework's robustness in scenarios with renewable penetration where stability margins are typically reduced.

\section{Conclusion and Future Work}\label{sec:conclusion}

In conclusion, this paper presents an LLM-driven agentic framework that automates the complete transient stability assessment workflow, from scenario generation and simulation execution to neural network architecture design. The proposed framework bridges the gap between natural language interaction and power system simulation by enabling automated disturbance scenario generation through tight integration with the simulator. The implementation of Retrieval-Augmented Generation and Chain-of-Thought methodologies demonstrates enhanced accuracy in translating user requirements into executable simulation scripts. For model design, the LLM-driven neural network design pipeline automatically discovers compact architectures that achieve strong classification performance while maintaining real-time inference capability. The integration of performance-driven feedback mechanisms effectively addresses the challenge of iterative architecture refinement without extensive human intervention. Ablation studies confirm that each component—retrieval augmentation, reasoning enhancement, and feedback correction—contributes to the overall effectiveness of the framework. Case studies on the IEEE 39-bus system, including scenarios with significant renewable penetration, validate the scalability and practical applicability of the proposed approach.

Future research will aim to enhance domain-specific capabilities through fine-tuning on dedicated power system code libraries. Efforts will also focus on developing physics-informed neural architecture search that incorporates power system dynamics directly into the design process. The goal is to establish a human-machine collaborative paradigm for power system analysis, which could serve as an LLM-powered engine for broader applications including optimal power flow, fault analysis, and market operations.

\section*{Appendix}
\addcontentsline{toc}{section}{Appendix}

\begin{center}
  \centering
  \captionof{table}{Large language models and parameter settings used in this study}
  \label{tab:llm_specifications}
  \begin{tabular}{@{}l l@{}}
    \toprule
    \textbf{Model Name} & \textbf{Version/API} \\
    \midrule
    GPT-4 & gpt-4-0613 \\
    text-embedding-ada-002 & v2 \\
    \midrule
    \multicolumn{2}{l}{\textit{Configuration Parameters}} \\
    Temperature & 0.3--0.7 \\
    Max Tokens & 1,024--4,096 \\
    Top-p & 0.95 \\
    API Rate Limiting & 10 requests/min \\
    Retry Strategy & Exponential backoff (max 3 retries) \\
    \bottomrule
  \end{tabular}
\end{center}
\FloatBarrier

\begin{center}
  \centering
  \captionof{table}{Training hardware specifications}
  \label{tab:hardware_specifications}
  \begin{tabular}{@{}l l@{}}
    \toprule
    \textbf{Component} & \textbf{Specification} \\
    \midrule
    GPU & NVIDIA RTX 4090 (24GB VRAM) \\
    CPU & Intel Core i9-13900K (24 cores, 32 threads) \\
    RAM & 64GB DDR5-5600 \\
    Storage & 2TB NVMe SSD (PCIe 4.0) \\
    \bottomrule
  \end{tabular}
\end{center}
\FloatBarrier

\subsection*{Prompt Examples}

\subsubsection*{Error Feedback Prompts Example}

\begin{tcolorbox}[top=10pt, colback=white, colframe=black, colbacktitle=black, center, enhanced, breakable,
attach boxed title to top left={yshift=-0.1in,xshift=0.15in}, boxed title style={boxrule=0pt,colframe=white,}, title=Error Feedback Prompts Example]
{\bf Performance Metrics:} Simulation execution failed with the following metrics:
\begin{itemize}
    \item Error Type: Runtime Error
    \item Error Message: ``NameError: name 'bus\_voltage' is not defined at line 42''
    \item Execution Time: 2.3 seconds before failure
    \item Code Segment: Wind farm integration scenario for IEEE 39-bus system
\end{itemize}

{\bf Problem Diagnosis:} The LLM-generated code references an undefined variable `bus\_voltage'. This indicates:
\begin{itemize}
    \item Variable initialization is missing in the code
    \item The Simulation Tools API method for retrieving bus voltage was not correctly invoked
    \item Parameter naming does not match Simulation Tools documentation conventions
\end{itemize}

{\bf Natural Language Feedback:} ``Your code attempted to access a bus voltage variable that was never initialized. In the Simulation Tools, bus voltage data must be retrieved using the correct API method after simulation completion. Please review the Simulation Tools documentation for voltage retrieval methods and ensure proper variable initialization before use.''

{\bf Optimization Strategy:}
\begin{itemize}
    \item Add proper variable initialization: `bus\_voltage = ss.Bus.v.v'
    \item Ensure simulation has completed before accessing results
    \item Use Simulation Tools-specific naming conventions for variables
\end{itemize}

{\bf Improvement Recommendations:}
\begin{enumerate}
    \item Modify line 42 to use correct Simulation Tools API: `bus\_voltage = ss.Bus.v.v' instead of direct variable access
    \item Add error handling with try-except blocks for robustness
    \item Include simulation completion check before result retrieval
    \item Reference Simulation Tools documentation for proper API usage patterns
\end{enumerate}
\end{tcolorbox}

\subsubsection*{Performance Feedback Example}

\begin{tcolorbox}[top=10pt, colback=white, colframe=black, colbacktitle=black, center, enhanced, breakable,
attach boxed title to top left={yshift=-0.1in,xshift=0.15in}, boxed title style={boxrule=0pt,colframe=white,}, title=Performance Feedback Example]
{\bf Performance Analysis:} The current model achieves 78.5\% accuracy on the validation set, with the following issues:
1. Training loss continues to decrease but validation loss begins to rise, indicating overfitting
2. Recall rate for unstable samples is low (65\%), potentially requiring enhanced learning for minority classes
3. Inference time is 15ms, exceeding real-time application requirements (<10ms)

{\bf Improvement Recommendations:}
1. Add Dropout layers (suggested dropout=0.3) and L2 regularization ($\lambda=1e-4$)
2. Use Focal Loss instead of cross-entropy loss, $\alpha=0.75$, $\gamma=2.0$
3. Reduce network depth or use knowledge distillation techniques to compress the model
4. Increase data augmentation strategies, particularly for unstable samples
\end{tcolorbox}

\subsubsection*{LLM-Simulation Prompt Example 1: Three-Phase Fault Scenario}

\begin{center}
\begin{tcolorbox}[top=10pt, colback=white, colframe=black, colbacktitle=black, center, enhanced, breakable,
attach boxed title to top left={yshift=-0.1in,xshift=0.15in}, boxed title style={boxrule=0pt,colframe=white,}, title=LLM-Simulation Prompt Example 1: Three-Phase Fault Scenario]

{\bf System Instruction:} {\small You are an expert power system engineer specializing in transient stability analysis. You will generate realistic fault scenarios for the IEEE 39-bus system using the ANDES simulator. Each scenario must include physically valid parameters and follow power system engineering principles.}

{\bf User Query:} {\small Generate a three-phase short circuit fault scenario on the IEEE 39-bus system with the following constraints:
\begin{itemize}
  \item Fault location: Bus 16 (critical hub bus)
  \item Fault inception time: 1.0 second
  \item Fault clearing time: 1.1 seconds (100 ms clearing duration)
  \item System loading: 100\% nominal load
  \item Fault impedance: $R_f = 0.01\Omega$, $X_f = 0.001\Omega$
  \item Expected outcome: Stable (rotor angle swing $< 180^\circ$)
\end{itemize}
Please provide: (1) ANDES simulation configuration; (2) Expected stability classification; (3) Physical reasoning.}

{\bf Assistant Response:} {\small

\textbf{ANDES Configuration:}
\begin{lstlisting}
fault_type = "3ph"  # Three-phase short circuit
fault_bus = 16      # Critical hub bus
t_fault = 1.0       # Fault inception (seconds)
t_clear = 1.1       # Fault clearing (seconds)
R_fault = 0.01      # Fault resistance (Ohms)
X_fault = 0.001     # Fault reactance (Ohms)
load_level = 1.0    # 100% nominal loading
\end{lstlisting}

\textbf{Stability Assessment:} This scenario is expected to result in \textbf{STABLE} classification because: (1) Bus 16 is a strong hub with multiple generator support; (2) 100 ms clearing time is within typical protection coordination; (3) Low fault impedance ensures fast voltage recovery post-clearing.

\textbf{Physical Reasoning:} The rotor angle deviation is expected to remain below $180^\circ$, indicating transient stability is maintained. Generator 10 (closest to Bus 16) will experience maximum angle swing but should recover within 5 seconds.}

\end{tcolorbox}
\end{center}

\subsubsection*{LLM-Simulation Prompt Example 2: Renewable Energy Integration (IEEE 39-bus)}

\begin{center}
\begin{tcolorbox}[top=10pt, colback=white, colframe=black, colbacktitle=black, center, enhanced, breakable,
attach boxed title to top left={yshift=-0.1in,xshift=0.15in}, boxed title style={boxrule=0pt,colframe=white,}, title=LLM-Simulation Prompt Example 2: Renewable Energy Integration (IEEE 39-bus)]

{\bf System Instruction:} {\small You are an expert in renewable energy integration and power system dynamics. Generate fault scenarios for the IEEE 39-bus system with integrated wind generation. Ensure the scenario respects renewable energy control characteristics and grid code requirements.}

{\bf User Query:} {\small Generate a three-phase fault scenario with 30\% wind penetration on the IEEE 39-bus system:
\begin{itemize}
  \item Base system: IEEE 39-bus
  \item Wind penetration: 30\% of total generation capacity
  \item Wind injection buses: Multiple strategic locations
  \item Wind control model: REGCV2 (virtual synchronous machine)
  \item Fault type: 3-phase short circuit on Bus 16
  \item Fault clearing: 150 ms
  \item Expected outcome: Angle + Frequency instability
\end{itemize}
Provide: (1) Modified system configuration; (2) Wind generator parameters; (3) Stability prediction.}

{\bf Assistant Response:} {\small

\textbf{System Configuration:}
\begin{lstlisting}
wind_penetration = 0.30  # 30% of total capacity
wind_buses = [16, 17, 21, 23]  # Strategic locations
wind_capacity_per_bus = 200  # MVA each
wind_control = "REGCV2"  # Virtual sync machine
fault_type = "3ph"
fault_bus = 16
t_clear = 0.15  # 150 ms
\end{lstlisting}

\textbf{Wind Generator Parameters:} Each wind generator (200 MVA) operates with REGCV2 control providing virtual inertia (H = 2.5s) and damping. During the 3-phase fault on Bus 16, wind generators will experience voltage dip and may temporarily reduce output due to low-voltage ride-through (LVRT) requirements.

\textbf{Stability Prediction:} This scenario is expected to exhibit \textbf{ANGLE + FREQUENCY INSTABILITY} because: (1) Wind generators reduce inertia support compared to synchronous generators; (2) Bus 16 is a critical transmission node with significant load; (3) 150 ms clearing time combined with 30\% wind penetration creates challenging dynamics. Rotor angle swings may exceed $120^\circ$ and frequency deviations may reach 2--3 Hz.}

\end{tcolorbox}
\end{center}

\printcredits

\bibliographystyle{cas-model2-names}

\bibliography{bib}

\end{document}